\documentclass[%
 aip,
 amsmath,amssymb,
 citeautoscript,
 reprint,%
]{revtex4-1}

\usepackage{graphicx}
\usepackage{dcolumn}
\usepackage{bm}

\usepackage[utf8]{inputenc}
\usepackage[T1]{fontenc}
\usepackage{mathptmx}
\usepackage{etoolbox}
\usepackage{xr-hyper}
\usepackage{hyperref}
\usepackage{comment}
\usepackage{xcolor}
\usepackage{float}

\makeatletter
\def\@email#1#2{%
 \endgroup
 \patchcmd{\titleblock@produce}
  {\frontmatter@RRAPformat}
  {\frontmatter@RRAPformat{\produce@RRAP{*#1\href{mailto:#2}{#2}}}\frontmatter@RRAPformat}
  {}{}
}%

\begin{document}
\preprint{AIP/123-QED}

\title{Strain effects on $n$-type doping in AlN}
\author{Haochen Wang}

\author{Chris G. Van de Walle}%
 \email{vandewalle@mrl.ucsb.edu}
\affiliation{ 
Materials Department, University of California Santa Barbara, Santa Barbara, CA 93106, USA
}%

\date{\today}

\begin{abstract}
Controllable doping in AlN and its alloys is essential for deep-ultraviolet light sources. 
Ionization energies for donors in AlN ($\mathrm{Si_{Al}}$, $\mathrm{S_N}$, $\mathrm{Se_N}$) are high.
We report first-principles calculations demonstrating that strain engineering can result in a reduction in ionization energies.
The donor levels for $\mathrm{S_N}$ and $\mathrm{Se_N}$ shift closer to the conduction-band minimum (CBM) under in-plane tensile strains,
driven by a downward shift of the CBM.
The most widely used donor, $\mathrm{Si_{Al}}$, forms a $DX$ center in AlN. We find that a 2.5\% in-plane tensile strain (which would be induced by pseudomorphic growth on GaN in experiment) shifts the ($+/-$) transition level from 271 meV to 98 meV below the CBM, which would enhance the electron concentration by three orders of magnitude. 
These results demonstrate that strain engineering offers an effective route to enhance doping levels in AlN.
\end{abstract}

\maketitle

Aluminum nitride (AlN) and aluminum gallium nitride (AlGaN) are strong candidates for deep-UV optoelectronic devices due to their ultrawide direct bandgap, mobility, and thermal conductivity. \cite{hickman_next_2021,kneissl_emergence_2019,tsao_ultrawidebandgap_2018,mishra_algangan_2002}
However, these devices require sufficiently high carrier concentrations to enable good transport to the active layers of light-emitting diodes and laser diodes. \cite{harris_compensation_2018, breckenridge_high_2021, taniyasu_electrical_2004}
Achieving highly conducting $n$-type AlN or Al-rich AlGaN remains a significant challenge.
Silicon is an excellent shallow donor in GaN, but it becomes a $DX$ center as the Al content increases in AlGaN.\cite{gotz_activation_1996,polyakov_properties_1998,Lyons_shallow_2012} 
$DX$ centers occur when an impurity expected to act as a shallow donor instead undergoes a large, bond-rupturing displacement, and traps electrons to become a deep acceptor.\cite{mooney_DX_1990}  
Previous studies on $\mathrm{Si_{Al}}$ in AlN reported a ($+/-$) charge-state transition level for the $DX$ center located in a range between 140 and 250~meV below the conduction-band minimum (CBM).\cite{gordon_hybrid_2014,harris_compensation_2018,Son_shallow_2011,collazo_progress_2011}
Other donors such as $\mathrm{O_N}$, $\mathrm{Ge_{Al}}$, and $\mathrm{C_{Al}}$ also form $DX$ centers in AlN, with even deeper transition levels.\cite{gordon_hybrid_2014,mccluskey_metastability_1998,lyons_effects_2014}
Chalcogen impurities on the nitrogen site have also been explored. They do not exhibit $DX$ behavior,\cite{gordon_sulfur_2015}
but the ($+/0$) transition level lies well below the CBM.\cite{lyons_deep_2025}

Strain can affect the properties of defects and impurities in semiconductors, and 
hydrostatic pressure has proven to be a valuable experimental tool.\cite{Gorczyca_HP_2024,Suski_pressure_1999}
Studies have shown that oxygen in GaN undergoes a transition from a shallow hydrogenic donor to a highly localized deep state above 20 GPa, whereas silicon remains hydrogenic up to at least 25 GPa.\cite{Wetzel_pressure_1997}
In $\mathrm{Al}_x \mathrm{Ga}_{1-x} \mathrm{N}$ alloys, hydrostatic pressure was found to shift the localized Si donor level downward relative to the CBM. \cite{Skierbiszewski_evidence_1999}
The evidence consistently indicates that hydrostatic compression leads to a lowering of the doping efficiency, leading to the expectation that \textit{expansion} could achieve a beneficial effect.
Hydrostatic expansion can of course not be implemented experimentally; however, biaxial tensile stress can readily be achieved in pseudomorphic growth.
In this paper, we explore 
the effects of biaxial strain on donors in AlN. 


Using first-principles density functional theory with a hybrid functional, 
we find that the ($+/-$) transition level of $\mathrm{Si_{Al}}$ and the ($+/0$) levels of $\mathrm{S_N}$ and $\mathrm{Se_{N}}$ move closer to the CBM under in-plane tensile strain (with ``in-plane'' referring to directions perpendicular to the $c$ axis).
At 2.5\% tensile strain, corresponding to the strain induced by pseudomorphic growth of AlN on GaN in experiment,\cite{Moram_2009_latt_para} the ($+/-$) level of $\mathrm{Si_{Al}}$ moves to 98 meV below the CBM, which could enhance the electron concentration by three orders of magnitude.
The $(+/0)$ level of $\mathrm{S_{N}}$ ($\mathrm{Se_{N}}$) shifts to 216 (294) meV below the CBM, which also enhances the electron concentration by three (four) orders of magnitude.
We will show that this behavior is primarily driven by the deformation-potential-induced lowering of the CBM under tensile strain.

Our calculations are based on generalized Kohn-Sham theory \cite{kohn_self-consistent_1965} and the Heyd-Scuseria-Ernzerhof (HSE) hybrid functional \cite{heyd_erratum_2006}implemented with projector augmented wave (PAW) potentials \cite{blochl_projector_1994,kresse_ultrasoft_1999} in the VASP code. \cite{kresse_efficient_1996}
We use a plane-wave cutoff energy of 500~eV, and 
the mixing parameter (33\%) is fitted to reproduce the experimental bandgap of AlN; we obtain 6.17~eV.
Our calculated lattice parameters $a=3.10$ {\AA}, $c=4.95$ {\AA} are in good agreement with experiment. \cite{Moram_2009_latt_para}
To investigate impurities, we employ a 288-atom wurtzite supercell, constructed as a $3\times 4 \times3$ multiple of the 8-atom orthorhombic unit cell.
Supercell calculations are performed using a single k-point ($\Gamma$). 
The incorporation of an impurity in a crystal is determined by its formation energy. 
For a Si impurity in charge state $q$ substituting on an Al site in AlN, the formation energy is given by:\cite{freysoldt_first-principles_2014}
\begin{equation}\label{eq:formation_energy}
E^f(\mathrm{Si}_{\mathrm{Al}}^q)=E_t (\mathrm{Si}_{\mathrm{Al}}^q)-E_t(\text {bulk})+\mu_{\mathrm{Al}}-\mu_{\mathrm{Si}}+q E_{\mathrm{F}}+\Delta^q
\end{equation}
where $E^f(\mathrm{Si}_\mathrm{Al}^q)$ is the formation energy of the defect in charge state $q$, $E_t(\mathrm{Si}_\mathrm{Al}^q)$ and $E_t(\text{bulk})$ are the total energies of the impurity-containing and pristine AlN supercells, and $\mu_\mathrm{Al}$ and $\mu_\mathrm{Si}$ are the chemical potentials of Al and Si.
The term $qE_\mathrm{F}$ accounts for the exchange of electrons with the Fermi level $E_\mathrm{F}$, which is referenced to the valence-band maximum (VBM).
$\Delta^q$ is a correction term accounting for the interactions of a charged supercell with its periodic images.\cite{freysoldt_fully_2009,freysoldt_electrostatic_2011}
The chemical potentials $\mu_{\text{Al}}$ and $\mu_{\text{N}}$ can vary within a range constrained by the enthalpy of formation of AlN, calculated to be $-$3.23 eV.
The chemical potentials of $\text{Si}$, $\text{S}$, and $\text{Se}$ are referenced to their elemental phases, and their upper bounds were determined by thermodynamic stability constraints by the secondary phases $\text{Si}_3 \text{N}_4$, $\text{Al}_2 \text{S}_3$ and $\text{Al}_2 \text{Se}_3$.
By evaluating formation energies of different charge states, the charge-state transition levels are obtained.
When applying in-plane tensile strain in the $c$ plane of AlN we allow relaxation along the $c$ direction.

Our calculated formation energies (see Fig.~S1 of the Supplementary Material) show that in the absence of strain
$\mathrm{Si_{Al}}$ forms a $DX$ center, leading to self-compensation of the shallow donor through the formation of an acceptor in the negative charge state.
The ($+/-$) transition level occurs at 271 meV below the CBM, somewhat deeper than previously calculated results. \cite{gordon_hybrid_2014,harris_compensation_2018}
In contrast, for $\mathrm{S_N}$ and $\mathrm{Se_N}$ the neutral charge state is also stable in the bandgap, in contrast to the results in Ref.~\onlinecite{gordon_sulfur_2015} but consistent with more recent calculations by Lyons {\it et al.}~\cite{lyons_deep_2025}; the ($+/0$) transition levels occur at 504 and 609 meV below the CBM, and the $(0/-)$ transition levels at 159 and 264 meV below the CBM (Fig.~S1 of the Supplementary Material).


\begin{figure}
\includegraphics[width=0.5\textwidth]{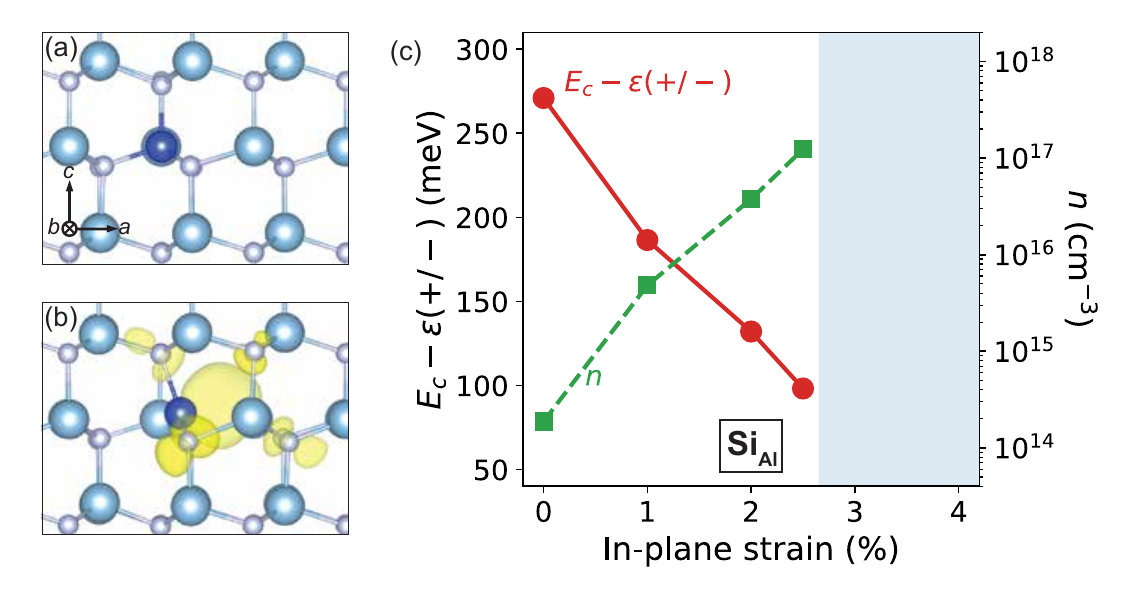}
\caption{\label{fig:Si_Al} 
Atomic structures of (a) $\mathrm{Si_{Al}^{+}}$ at the substitutional site and (b) $\mathrm{Si_{Al}^{-}}$ in the most stable $DX$ configuration in unstrained AlN.
The Si atom is shown as a dark blue sphere, Al atoms as light blue spheres, and N atoms as small white spheres.
The yellow lobes represent the charge density associated with the occupied $DX$ state within the bandgap, with isosurfaces set to 5\% of the maximum value.
(c) Ionization energy of $\mathrm{Si_{Al}}$ in AlN and electron concentration $n$ as a function of in-plane tensile strain.
The shaded area indicates the strain regime where $\mathrm{Si_{Al}^-}$ can assume a substitutional configuration. 
}
\end{figure}

The atomic configuration of 
$\mathrm{Si_{Al}^{+}}$ is shown in Fig.~\ref{fig:Si_Al}(a): the Si atom occupies an Al substitutional site, leading to a 6\% contraction of the surrounding Si-N bonds (1.77 {\AA}), referenced to the bulk Al-N bond length (1.89 {\AA}).
The difference between axial and in-plane Al-N bond lengths is less than 0.6\% (we use the term ``in-plane'' to refer to the three bonds that are not along the $c$ axis).
The local configuration of the neutral $\mathrm{Si_{Al}^0}$ is the same as for $\mathrm{Si_{Al}^{+}}$, indicating that the extra electron 
is delocalized in the conduction band.
In contrast, for $\mathrm{Si_{Al}^{-}}$ a $DX$ configuration is formed as shown in Fig.~\ref{fig:Si_Al}(b):
the Si atom shifts off the substitutional site, resulting in one broken in-plane Si–N bond elongated by 23\%, while the remaining three Si–N bonds contract by 4\%.
Figure~\ref{fig:Si_Al}(b) also shows the charge density of the $DX$-induced occupied state in the gap, illustrating its localized and broken-bond character.
We also identify a metastable $DX$ configuration for $\mathrm{Si_{Al}^{-}}$, which was reported as the lowest-energy configuration in Ref.~\onlinecite{gordon_hybrid_2014}, characterized by the breaking of the axial Si–N bond along the $c$ axis; in our calculations this configuration lies 200~meV above the most stable structure shown in Fig.~\ref{fig:Si_Al}(b).

We now systematically investigate the impact of in-plane tensile strain.
Figure~\ref{fig:Si_Al}(c) shows the ionization energy of $\mathrm{Si_{Al}}$ in AlN [the energy difference between the ($+/-$) transition level and the CBM].
As strain increases, ($+/-$) moves closer to the CBM;
at 2.5\% strain (which would be induced by pseudomorphic growth on GaN in experiment), the ionization energy is less than 100~meV, which allows efficient thermal activation.
Beyond 2.5\% strain, our calculations indicate the appearance of an alternative stabilized configuration for $\mathrm{Si_{Al}^-}$, structurally identical to the substitutional configuration of $\mathrm{Si_{Al}^+}$ and $\mathrm{Si_{Al}^0}$. 
In this configuration the two electrons are delocalized, and hence the total energy cannot be reliably assessed within our finite-size supercell. 
We can therefore not conclusively assess which of the two possible configurations of $\mathrm{Si_{Al}^-}$ is lower in energy: the localized $DX$ configuration or the delocalized substitutional configuration. 
Either way, our results show a clear trend toward shallow-donor behavior under tensile strain.
In Fig.~\ref{fig:Si_Al}(c), we use shading to represent the strain regime where the delocalized substitutional configuration of $\mathrm{Si_{Al}^-}$ appears.

The position of the transition level determines the electron concentration $n$ in semiconductor devices, as shown in Sec.~B of Supplementary Material.
Figure~\ref{fig:Si_Al}(c) shows the calculated electron concentration $n$ in \text{Si}-doped AlN as a function of in-plane tensile strain. 
Here we assume $N_\mathrm{Si} = 10^{18}\ \mathrm{cm^{-3}}$, a typical value. 
For the unstrained case, the $(+/-)$ transition level is 271 meV below the CBM, resulting in $n=1.9 \times10^{14} \ \mathrm{cm^{-3}}$.
The strain-induced reduction of the ionization energy significantly enhances the electron concentration. 
At 2.5\% tensile strain, where the $(+/-)$ level is at 98 meV below the CBM, $n=1.2\times10^{17}\ \mathrm{cm^{-3}}$, three orders of magnitude higher than in the unstrained material.

\begin{figure}
\includegraphics[width=0.5\textwidth]{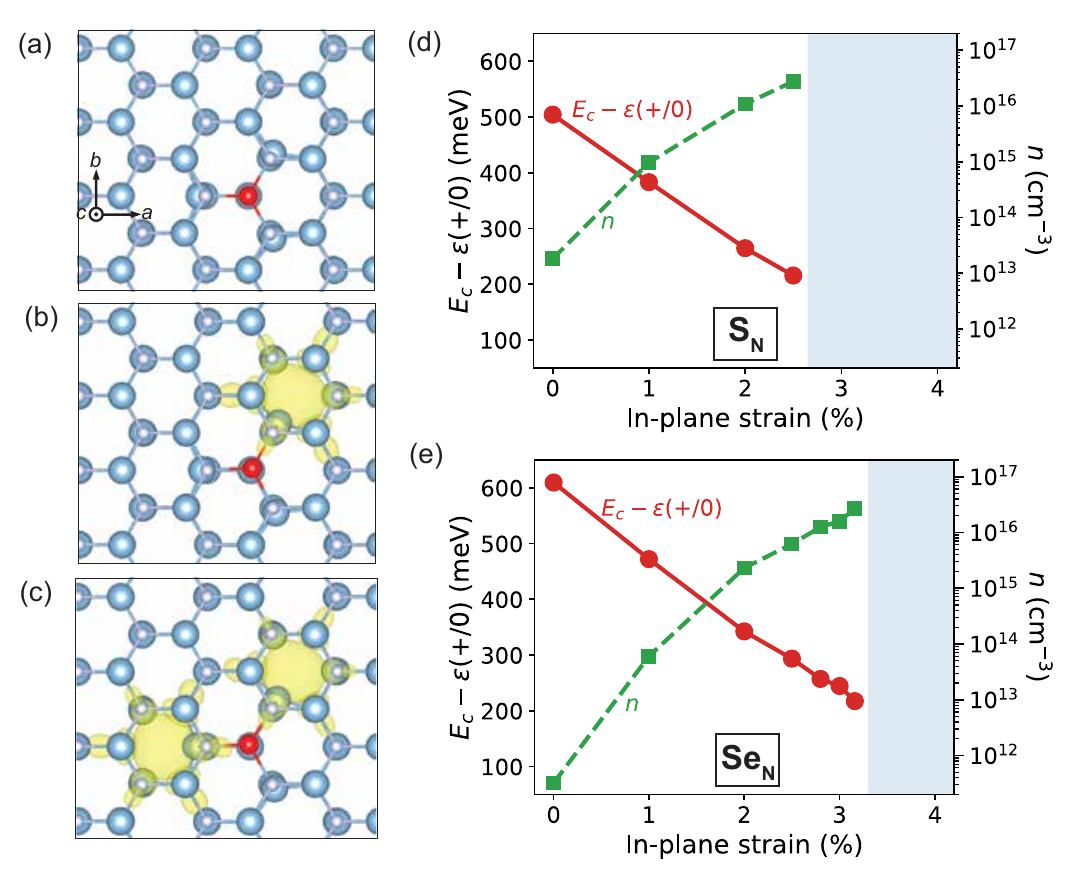}
\caption{\label{fig:S_N} 
Atomic structures of (a) $\mathrm{S_{N}^{+}}$, (b) $\mathrm{S_{N}^{0}}$ and (c) $\mathrm{S_{N}^{-}}$ in unstrained AlN.
The sulfur atom is shown as red sphere.
The yellow lobes in (b) and (c) represent the charge density associated with the occupied impurity states in the bandgap, with isosurfaces set to 5\% of the maximum value.
(d)-(e) Ionization energy and electron concentration $n$ as a function of in-plane tensile strain for (d) $\mathrm{S_{N}}$ and (e) $\mathrm{Se_{N}}$ in AlN.
The shaded areas indicate the appearance of a delocalized substitutional configuration of neutral and negative charged states of $\mathrm{S_{N}}$ and $\mathrm{Se_{N}}$.
}
\end{figure}

We now turn to $\mathrm{S_N}$, for which the atomic configurations of $\mathrm{S_N^+}$, $\mathrm{S_N^0}$ and $\mathrm{S_N^{-1}}$ are shown in Figs.~\ref{fig:S_N}(a)--(c).
In the positive charge state, the S atom resides on the N site, with the surrounding Al--S bonds expanding by 17\% (in-plane) and 20\% (axial).
In the neutral state, the S atom slightly shifts off the substitutional site; one in-plane Al neighbor is pushed further away with the bond elongated by 22\%.
The charge density associated with the occupied state in the bandgap is shown in Fig.~\ref{fig:S_N}(b):
The electron is localized within a contracted hexagonal cavity, consistent with the observations in Ref.~\onlinecite{lyons_deep_2025}.
In the negative charge state, two contracted cavities localize the electrons [Fig.~\ref{fig:S_N}(c)].

Figure~\ref{fig:S_N}(d) shows the ionization energy of $\mathrm{S_N}$ in AlN and the electron concentration $n$ as a function of in-plane tensile strains, for a dopant concentration of $10^{18}\ \mathrm{cm^{-3}}$.
As strain increases, the ($+/0$) transition level moves closer to the CBM, reaching 265 meV below the CBM at 2.5\% strain. This enhances the electron concentration by three orders of magnitude, from $1.8 \times10^{13} \ \mathrm{cm^{-3}}$ in the unstrained case to $2.7 \times10^{16} \ \mathrm{cm^{-3}}$ under 2.5\% strain.
Beyond 2.5\% strain, an alternative stable configuration for $\mathrm{S_{N}^0}$ and $\mathrm{S_{N}^-}$ appears. 
For instance, at 2.8\% strain for both $\mathrm{S_N^0}$ and $\mathrm{S_N^-}$, all three in-plane S–Al bonds around the sulfur atom become equivalent (2.24~{\AA}, uniformly expanded by 17\%), identical to the configuration of $\mathrm{S_N^+}$. 
In this configuration, $\mathrm{S_N^0}$ and $\mathrm{S_N^-}$ look like $\mathrm{S_N^+}$ surrounded by one or two delocalized electrons.
Because of this delocalized character, the relative stability between this configuration and the localized configuration cannot be accurately resolved in our finite-size supercells.
Nevertheless, 
the decrease in the donor ionization energy under tensile strain supports efficient dopant activation at room temperature.
In Fig.~\ref{fig:S_N}(d), we use shading to represent the strain regime where the substitutional configuration of $\mathrm{S_N^0}$ and $\mathrm{S_N^-}$ appears.

The atomic structures of $\mathrm{Se_N}$ are similar to $\mathrm{S_N}$. 
The ionization energy of $\mathrm{Se_N}$ and electron concentration in AlN as a function of in-plane tensile strains are shown in Fig.~\ref{fig:S_N}(e); we again assume a dopant concentration of $10^{18}\ \mathrm{cm^{-3}}$.
For $\mathrm{Se_N}$, the delocalized substitutional configuration of $\mathrm{Se_N^0}$ and $\mathrm{Se_N^-}$ appears when strain exceeds 3.2\% (shading in Fig.~\ref{fig:S_N}(e)).
The ionization energy decreases with increasing tensile strain, from 609 meV at zero strain to 294 meV at 2.5\% strain (which would be induced by pseudomorphic growth of AlN on GaN in experiment).
The reduction of ionization energy enhances the electron concentration by four orders of magnitude, from $3.2 \times10^{11} \ \mathrm{cm^{-3}}$ at zero strain to $6.2 \times10^{15} \ \mathrm{cm^{-3}}$ at 2.5\% strain.

\begin{figure}
\includegraphics[width=0.4\textwidth]{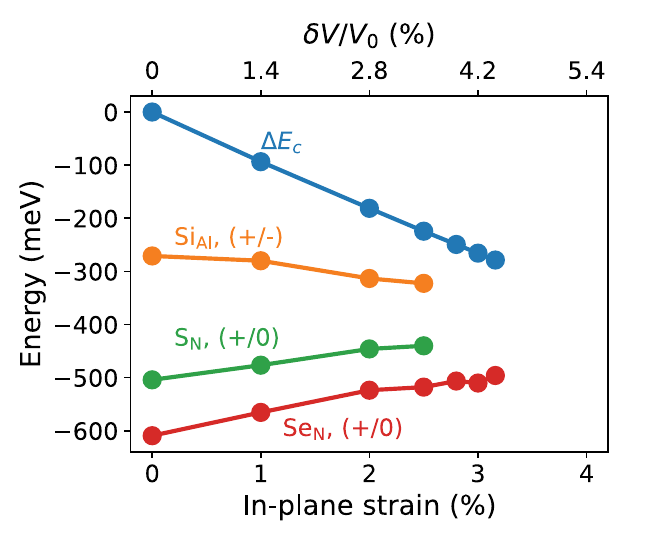}
\caption{\label{fig:absolute_ene} 
($+/-$) of $\mathrm{Si_{Al}}$, ($+/0$) of $\mathrm{S_N}$ and $\mathrm{Se_N}$, and the CBM energy $E_c$ as a function of strain shown on an absolute energy scale.
The top axis shows the fractional volume change under corresponding strain.
}
\end{figure}

We now discuss the physical origin of the enhanced dopant activation.
The change in ionization energy under strain is determined by two factors:
(1) the shift of the CBM on an absolute energy scale, and
(2) the relative change in energy of the impurity in different charge states.
In the case of $\mathrm{Si_{Al}}$, 
the transition level $(+/-)$ corresponds to the Fermi-level position for which 
$E^f(\mathrm{Si}^+_\mathrm{Al})=E^f(\mathrm{Si}^-_\mathrm{Al})$, for which Eq.~(\ref{eq:formation_energy}) yields $\varepsilon(+/-)=\frac{1}{2}[E_t(\mathrm{Si}^-_\mathrm{Al})-E_t(\mathrm{Si}^+_\mathrm{Al})+\Delta^--\Delta^+]$.
%
In Fig.~\ref{fig:Si_Al}(d) we separate the contributions of these effects by plotting the CBM energy $E_c$ and the transition levels
as a function of strain on an absolute energy scale, referenced to the CBM in unstrained material.
The shift of the CBM is governed by  $\Delta E_c=a_c \delta V/V_0$ where $\delta V/V_0$ is the fractional volume change and $a_c$ is the absolute deformation potential for the CBM ($a_c=-6.4 \ \text{eV}$ from Ref.~\onlinecite{janotti_absolute_2007}).
Figure~\ref{fig:Si_Al}(d) shows that the downward shift of the CBM is the dominant factor responsible for the reduction in ionization energy of $\mathrm{Si_{Al}}$: the ($+/-$) transition level of $\mathrm{Si_{Al}}$ remains almost constant, changing by only 51~meV over the range of strain up to 2.5\%, while the CBM is lowered by 224 meV.
For $\mathrm{S_N}$ and $\mathrm{Se_N}$, Fig.~\ref{fig:absolute_ene} shows that their ($+/0$) levels shift slightly upwards, indicating that in the neutral charge state the energy of the localized electron increases when the interstitial space becomes more strained.


In summary, we have investigated the effects of strain on donors 
in AlN.
Applying in-plane tensile strain shifts the transition levels closer to the CBM, 
an effect that is primarily driven by the downward shift of the CBM.
In particular, for $\mathrm{Si_{Al}}$, a 2.5\% strain lowers the $(+/-)$ transition level to 98~meV below the CBM, resulting in a three-orders-of-magnitude increase in the electron concentration $n$. 
For $\mathrm{S_N}$, the $(+/0)$ level shifts to 216~meV below the CBM at 2.5\% strain, also leading to a three-orders-of-magnitude enhancement in $n$. 
For $\mathrm{Se_N}$, at 2.5\% strain the $(+/0)$ level moves to 294~meV below the CBM and the electron concentration increases by approximately four orders of magnitude.
These findings demonstrate that strain engineering offers an effective approach to enhance $n$-type doping efficiency in AlN.
\\

See the Supplementary Material for formation energy diagrams of $\mathrm{Si_{Al}}$, $\mathrm{S_N}$, and $\mathrm{Se_N}$ in unstrained AlN; and the calculation of carrier concentration.

\begin{acknowledgments}
    This work is supported by 
    SUPREME, one of seven centres in JUMP 2.0, a Semiconductor Research Corporation program sponsored by the Defense Advanced Research Projects Agency. 
    The work used Stampede3 at Texas Advanced Computing Center (TACC) through allocation DMR070069 from the Advanced Cyberinfrastructure Coordination Ecosystem: Services \& Support (ACCESS) program, which is supported by National Science Foundation grant No. 2138259, 2138286, 2138307, 2137603 and 2138296. This research also used resources of the National Energy Research Scientific Computing Center, a DOE Office of Science User Facility supported by the Office of Science of the U.S. Department of Energy under Contract No. DE-AC02-05CH11231 using NERSC award BES-ERCAP0028497.
\end{acknowledgments}

\section*{AUTHOR DECLARATIONS}
\subsection*{Conflict of Interest}
The authors have no conflicts to disclose.

\section*{Data Availability Statement}
The data that support the findings of this study are available from the corresponding authors upon reasonable request.

\bibliography{main}

\end{document}